# Correlations of Amino Acids with Secondary Structure Types: Connection with Amino Acid Structure

Saša Malkov,[1] Miodrag V. Živković,[1] Miloš V. Beljanski,[2] Snežana D. Zarić[3*]

[1] *Department of Mathematics, University of Belgrade, Belgrade, Serbia and Montenegro*
[2] *Institute of General and Physical Chemistry, Belgrade, Serbia and Montenegro*
[3] *Department of Chemistry, University of Belgrade, Belgrade, Serbia and Montenegro, szaric@chem.bg.ac.yu*

The correlations of primary and secondary structures were analyzed using proteins with known structure from Protein Data Bank. The correlation values of amino acid type and the eight secondary structure types at distant position were calculated for distances between -25 and 25. Shapes of the diagrams indicate that amino acids polarity and capability for hydrogen bonding have influence on the secondary structure at some distances. Clear preference of most of the amino acids towards certain secondary structure type classifies amino acids into four groups: α-helix admirers, strand admirers, turn and bend admirers and the others. Group four consists of *His* and *Cis*, the amino acids that do not show clear preference for any secondary structure. Amino acids from a group have similar physicochemical properties, and the same structural characteristics. The results suggest that amino acid preference for secondary structure type is based on the structural characteristics at Cβ and Cγ atoms of amino acid. α-helix admirers do not have polar heteroatoms on Cβ and Cγ atoms, nor branching or aromatic group on Cβ atom. Amino acids that have aromatic groups or branching on Cβ atom are strand admirers. Turn and bend admirers have polar heteroatom on Cβ or Cγ atoms or do not have Cβ atom at all. Our results indicate that polarity and capability for hydrogen bonding have influence on the secondary structure at some distance, and that amino acid preference for secondary structure is caused by structural properties at Cβ or Cγ atoms.

## Introduction

The prediction of protein structure from its sequence is fundamental problem and its solution is goal of all protein folding theories. There are many methods that consider protein folding (1-15) and many of them use some information about the protein secondary structure (1, 3, 8-16).

The principle that secondary structures of proteins are determined by their amino acid sequence is a basis of the protein secondary structure prediction methods. It is known for long time that different amino acids have distinct propensities for the adoption of helical, strand, and random coil conformation (17, 18-23). There are many secondary structure prediction algorithms that are based on these propensities (24). Correlation of amino acid and secondary structure is also important for *de novo* protein design.

Furthermore, amino acids propensities are position dependent. It is known that amino acids show distinct position-dependent helix-forming propensities near the ends of α-helices (25). It also was shown that the amino acids have very strong position-dependent propensities throughout the length of a helix. These propensities are connected to amino acids hydrophobicity (26, 27). In β-strands alternating hydrophobic/polar patterns are very frequent (28).

Although individual amino acids show intrinsic propensities towards certain secondary structure type (18-23), these preferences are modulated by the sequence segments within which they reside (29-31). Moreover, any amino acid in the protein could have influence on secondary structure type at certain position. It is practically impossible to consider the influence of the complete amino acid sequence ($R_1$, $R_2$, ... $R_n$) on the secondary structure type $S_j$ at the fixed position *j*. It is therefore necessary to limit the analysis to a local segment of the amino acid sequence. Local structural information is often contained in local parts of sequences (32, 33). It is estimated that local information contains roughly 65% of the secondary structure information (34). There are also approaches that consider non-local interactions in the sequence (35-38).

In order to check influence of local segment on the secondary structure type $S_j$ it is natural to consider the symmetrical window in the amino acid sequence centered at position *j*. The width of the window has to be chosen in the way that: (i) the influence of the rest of the sequence on $S_j$ could be neglected and (ii) the influence of amino acids from the window on $S_j$ could be effectively estimated. A possible approach to estimate influence of amino acids is to consider separately and independently the influences of particular nearby amino acids on $S_j$, assuming the approximation is good enough. Such an approach is typical (39).

Statistical approach to the protein secondary structure prediction is usually based on amino acids propensities to be part of different types of secondary structures (18). The propensity of the amino acid $A$ towards the secondary structure $S$ is the quotient of probability (relative frequency) of $A$ inside $S$ and the overall probability of $A$,

$$P_{A,S} = \frac{P(A|S)}{P(A)} = \frac{n_{A,S}/n_S}{n_A/n},$$

where $n_{A,S}$, $n_S$, $n_A$ and $n$ are the numbers of occurrences of $A$ as parts of $S$, occurrences of $S$, occurrences of $A$, and the sample size, respectively (40).

More advanced *information content* is based on propensities and defined by

$$I(S;A) = \log(P_{A,S}) - \log(P_{A,\neg S}) = \log\left(\frac{n_{A,S}/n_S}{n_{A,\neg S}/n_{\neg S}}\right),$$

(41). Here $\neg S$ denotes the appearance of "non $S$" secondary structure.

Correlations of amino acids with secondary structure types are used to predict protein structure, but they are also important for understanding the forces that stabilize protein structures. Increasing number of proteins with known structure, makes the correlations more reliable and better represents the interactions in proteins. Noncovalent interactions play important role in stabilizing tertiary structures of proteins and there are many types of noncovalent interactions that should be considered. The analysis of data from Protein Data Bank (PDB) (42) enabled to elucidate different kind of noncovalent interactions in proteins (43, 44). It was found out that new type of cation-π interactions, interactions of ligands coordinated to a metal with aromatic groups of amino acid residues, play role in stabilizing structure of proteins (43).

In this paper we describe the influence of amino acids on secondary structure types. Secondary structures types are α-helix (H), isolated β-bridge (B), extended strand (E), 3-helix (G), 5-helix (I), hydrogen bonded turn (T), and bend (S) (45). All other structural elements, not belonging to these secondary structure types, are considered *coil* and denoted by C. Secondary structure types are often reduced to only three; H, E, and C (24, 46). In this work we calculated influence of amino acids on all eight types, using statistical correlation data for proteins from PDB (42). Previously used information content (47) is useful for secondary structure prediction, because of its additivity. However, statistical correlation seems to be more natural characteristics of the relation between the primary and secondary structures. The values of correlation coefficients are determined in a local window range, considering influence of nearby amino acids on secondary structure. The statistical significances of computed corelations are ranked. The results enable to find amino acid preferences towards secondary structure types and to evaluate how far the dependence of secondary structure on amino acid is distributed across the chain. Based on the preferences all amino acids are classified in four groups. Amino acids in the same group have the same structural properties. To the best of our knowledge, these are the first results clearly connecting amino acid preferences with their structures.

## Method

Secondary structure types are assigned by DSSP (45). They are denoted using letters: H for α-helix, B for isolated β-bridge, E for extended strand, G for 3-helix, I for 5-helix, T for hydrogen bonded turn and S for bend. All other structural elements, not belonging to these secondary structure types, are considered *coil* and denoted by C. Here we consider all eight secondary structure types, including coils.

### Computational Model

Consider a set $P$ of $n$ protein chains. Primary structures of these protein chains are described by sequences $a_1, \ldots a_n$. If len($i$) denotes the length of the sequence $a_i$, then residues of the sequence $a_i$ are $a_{i,1}, \ldots a_{i,\text{len}(i)}$, $1 \leq i \leq n$. The corresponding assigned secondary structures are described by sequences $b_1, \ldots, b_n$, where $b_i$ is a sequence $b_{i,1}, \ldots, b_{i,\text{len}(i)}$, $1 \leq i \leq n$.

If $A$ is a logical expression, then the indicator variable $I(A)$ is defined by:

$$I(A) = \begin{cases} 1 & , \quad A = true \\ 0 & , \quad A = false \end{cases}$$

Let $X_{ij}(s)=I(b_{ij} = s)$ and $Y_{ij}(p)=I(a_{ij} = p)$ denote binary random variables corresponding to events that the secondary structure type assigned to residue $a_{ij}$ is $s$, and that $a_{ij}$ is the amino acid $p$, respectively. Consider the window in the sequence $i$, centered at the position $j$. Let

$$Z_{ij}(s, p, \tau) = X_{ij}(s) \cdot Y_{i,j+\tau}(p) \qquad (1)$$

denote the random variable corresponding to the joint event that the secondary structure type assigned to $a_{ij}$ (amino acid in the center of the window) is $s$ and that $a_{i,j+\tau}$ (amino acid at the offset $\tau$ from the window center) is $p$.

The offset range is determined by the size of the window. The definition of $Z_{ij}(s, p, \tau)$ is valid only if both $j$ and $j+\tau$ are between 1 and len($i$). Let $S(\tau)$ denote the set of valid pairs $(i,j)$:

$$S(\tau) = \{(i, j) \mid 1 \leq j, j + \tau \leq len(i)\}.$$

Let $T(\tau)$ denote the size of $S(\tau)$.

Let us now introduce some notation. $NPS(s, p, \tau)$ - the number of times the amino acid $p$ occurs at distance $\tau$ from the position $j$, where the secondary structure type is $s$:

$$NPS(s, p, \tau) = \sum_{S(\tau)} Z_{ij}(s, p, \tau);$$

$NP(p,\tau)$ - the number of occurrences of amino acid $p$ at positions $j+\tau$, such that the position $j$ is inside the same chain, for all secondary structure types $s$:

$$NP(p,\tau) = \sum_s NPS(s, p, \tau);$$

$NS(s,\tau)$ - the number of times the secondary structure type $s$ is assigned at the position $j$, such that the position $j+\tau$ is inside the same chain:

$$NS(s,\tau) = \sum_p NPS(p, s, \tau).$$

The total count of valid sample pairs at distance $\tau$ is

$$T(\tau) = \sum_{p,s} NPS(s, p, \tau). \quad (2)$$

The correlation coefficient of random variables $X$ and $Y$ is defined by

$$\rho(X,Y) = \frac{Cov(X,Y)}{\sqrt{Var(X) \cdot Var(Y)}},$$

(see (48), for example). If both variables are binary, then

$$\rho(X,Y) = \frac{\overline{XY} - \overline{X}\,\overline{Y}}{\sqrt{\overline{X}(1-\overline{X})\overline{Y}(1-\overline{Y})}}.$$

The correlation coefficient is always in the range [-1, 1]. It is 0 if $X$ and $Y$ are independent. The correlation coefficient is 1 or –1 if and only if the random variables are linearly dependent.

Consider the correlation of random variables $X_{ij}$ and $Y_{ik}$,

$$\rho(X_{ij}(s), Y_{ik}(p)) = \frac{\overline{Z_{ij}(s,p,\tau)} - \overline{X_{ij}(s)}\,\overline{Y_{ik}(p)}}{\sqrt{\overline{X_{ij}(s)}(1-\overline{X_{ij}(s)})\overline{Y_{ik}(p)}(1-\overline{Y_{ik}(p)})}},$$

where $k = j+\tau$ and $Z_{ij}(s,p,\tau)$ is defined by Equation 1. Assuming that distributions of $X_{ij}$ and $Y_{ik}$ depend on $p$, $s$ and $\tau$ only (i.e. they are independent on the sequence choice and the absolute position inside the sequence), we estimate the correlation coefficients at offset $\tau$ by

$$\rho(s,p,\tau) = \frac{\overline{Z(s,p,\tau)} - \overline{X(s,\tau)}\,\overline{Y(p,\tau)}}{\sqrt{\overline{X(s,\tau)}(1-\overline{X(s,\tau)})\overline{Y(p,\tau)}(1-\overline{Y(p,\tau)})}}.$$

The estimates of means of $X$, $Y$ and $Z$ are

$$\overline{X(s,\tau)} = NS(s,\tau)/T(\tau),$$
$$\overline{Y(p,\tau)} = NP(p,\tau)/T(\tau),$$
$$\overline{Z(s,p,\tau)} = NPS(s,p,\tau)/T(\tau).$$

Hence the correlation coefficient estimate is

$$\rho(s,p,\tau) =$$

$$\frac{NPS(s,p,\tau)T(\tau) - NP(p,\tau)NS(s,\tau)}{\sqrt{NP(p,\tau)(T(\tau)-NP(p,\tau))NS(s,\tau)(T(\tau)-NS(s,\tau))}} \quad (3)$$

The value of $\rho(s, p, \tau)$ is positive (negative, zero) if the pair $(p, s)$ occurs more (less, equally) frequently at the distance $\tau$ than it would occur if $p$ and $s$ were independent at the distance $\tau$.

In order to evaluate the significance of the correlation coefficient, we compute the statistics $t_s$

$$t_s = \rho\sqrt{\frac{n-2}{1-\rho^2}},$$

where $n=T(\tau)$. Under assumption that the correlation coefficient is $0$, the distribution of the statistics $t_s$ is $t$-distribution with $n-2$ degrees of freedom (see (48) for example).

If the sample size $n$ is large, then $t$-distribution is approximated by the normal distribution $N(0,1)$. Let the *null hypothesis* be that $X$ and $Y$ are independent, i.e. that there is no dependence of the secondary structure type $s$ and amino acid $p$ at distance $\tau$. The null hypothesis is considered false if it implies that the probability to obtain correlation coefficient estimate with absolute value greater than calculated is less than 0.001. If the null hypothesis is true, using the normal distribution approximation we obtain that the probability of the event "$|t_s|$ is greater than $t_{lim}$" is 0.05 for $t_{lim} = 1.96$. Hence, the correlation coefficient is significant, and we consider $X$ and $Y$ are dependent if $|t_s| \geq 1.96$. If we denote the corresponding value of the correlation coefficient by $\rho_{lim}$, then the correlation coefficient is significant if

$$|\rho| \geq \rho_{lim} = \frac{t_{lim}}{\sqrt{t_{lim}^2 + n - 2}} = \frac{1.96}{\sqrt{3.84 + n}}. \quad (4)$$

### Data Sets

As a source of protein data we used Protein Data Bank (PDB), release #103 from January 2003, containing 18482 proteins (42). The secondary structures assignment is performed by the program DSSP (45). There are many families of proteins that are overrepresented in PDB. The full set of protein sequences is filtered to eliminate redundant data - we used the PDBSELECT list of nonredundant protein chains (49), with the threshold 25%. The resulting set contains 1737 sequences with 282,329 amino acid residues. The sample size ranges from 244,329 (for $|\tau| = 25$) to 282,329 (for $\tau = 0$). The corresponding values of $\rho_{lim}$ are 0.0040 (for $|\tau| = 25$), and 0.0037 (for $\tau = 0$).

### Results and Discussion

The values of correlation of amino acids with secondary structure types are computed for the PDBSELECT subset of protein sequences using Equation 3. The 8160 correlation values are calculated (8 types of secondary structures, 20 amino acids and 51 offsets). Based on correlation values amino acids are classified in four groups according to their preferences to participate in secondary structures (Table I). The groups are: α-helix admirers, strand admirers, turn and bend admirers, and the fourth group consists of amino acids that do not show preference for any of secondary structure types. The correlation values are represented in Figures 1-4. Every Figure contains diagrams for one group of amino acids; there are separate diagrams for every amino acid. Diagrams consist of eight graphs, representing the correlation of the amino acid and eight secondary structure types. The diagrams show that almost every amino acid prefers one secondary structure type, since it has much higher correlation values for one secondary structure that for others (Figures 1-3).

**Table I - Values of correlation coefficients of amino acids and secondary structure types at same position, and elements of amino acids structure.** The table contains correlation values calculated using Formula 3, for distance $\tau = 0$, and multiplied by 10,000. To emphasize the most important correlations, significant positive correlation coefficients ($\rho > 0.015$) are presented in bold and significant negative correlation coefficients ($\rho < -0.015$) are presented in italic. Three rightmost columns contain information on structural properties of amino acids. If an amino acid has branching on Cβ, or aromatic Cγ atom, the appropriate cell is marked. Otherwise it is empty. If there is a polar heteroatom on Cβ or Cγ, the chemical symbol for the atom is presented in the last column.

(a) There is a nonpolar sulfur atom on Cγ atom in *Met*.

| | | α-helix (H) | Strand (E) | Turn (T) | Bend (S) | 3-helix (G) | Coil (C) | Branch on Cβ | Arom. Cγ | Polar hetero atom on Cβ or Cγ |
|---|---|---|---|---|---|---|---|---|---|---|
| α-helix admirers | Ala | **825** | *-357* | *-183* | *-279* | 90 | *-248* | | | |
| | Leu | **766** | **174** | *-476* | *-346* | -34 | *-413* | | | |
| | Glu | **642** | *-392* | 32 | -58 | 131 | *-358* | | | |
| | Gln | **413** | *-247* | -60 | -46 | 57 | *-159* | | | |
| | Arg | **297** | -107 | -122 | -53 | -1 | -106 | | | |
| | Met | **265** | -19 | *-212* | *-169* | 0.2 | 4 | | | (a) |
| | Lys | **242** | *-264* | 75 | 52 | 6 | -101 | | | |
| Strand admirers | Val | *-172* | **1280** | *-581* | *-350* | *-274* | *-281* | X | | |
| | Ile | 144 | **945** | *-566* | *-371* | *-223* | *-326* | X | | |
| | Tyr | -53 | **470** | *-192* | *-150* | 9 | *-184* | | X | |
| | Phe | 91 | **458** | *-256* | *-191* | -5 | *-163* | | X | |
| | Thr | *-391* | **281** | *-198* | 120 | *-161* | **287** | X | | O |
| | Trp | 87 | **157** | -146 | -123 | 66 | -89 | | X | |
| Turn and bend admirers | Gly | *-1050* | *-492* | **1380** | **846** | -112 | 73 | | | |
| | Asn | *-379* | *-450* | **567** | **302** | 24 | **217** | | | O |
| | Pro | *-895* | *-664* | **456** | 136 | **164** | **1160** | | | N |
| | Asp | *-304* | *-623* | **357** | **339** | 157 | **400** | | | O |
| | Ser | *-373* | *-203* | 61 | **249** | 156 | **334** | | | O |
| Other | Cys | -42 | 94 | -74 | -67 | -40 | 70 | | | S |
| | His | -118 | 2 | 25 | 66 | 61 | 34 | | | N |

**Amino Acids And Secondary Structure Types At Same Position**

Correlation values represented in Figures 1-4 characterize the behavior of different amino acids. In Table I the correlation values for the secondary structure type at the position of amino acid are presented. These numbers correspond to the values from diagrams at the position with $\tau = 0$. Amino acids in Table I are classified according to their preference to participate in specific secondary structures. Amino acids in each group are ordered by the correlation values. Our results are in some extent in agreement with calculated propensities of amino acids by Chou and Fasman (17), but there are also substantial differences, as it will be discussed for each group.

*Propensities Of Amino Acids*

**α-helix admirers.** The amino acids from the first group in Table I (*Ala, Leu, Glu, Gln, Arg, Met* and *Lys*) are *helix admirers*, showing preference to build α-helices. Because of the high level of their correlations with α-helices, three of them (*Ala, Leu, Glu*) could be further classified as *strong helix admirers*. The previous findings, (17, 50) that *Glu, Ala* and *Leu* are found most frequently in helical regions, are in agreement with our results. However, Chou and Fasman finding that *His* and *Val* are also very frequent differs from our results, where both of these amino acids show negative correlation with α-helix.

(1.a)
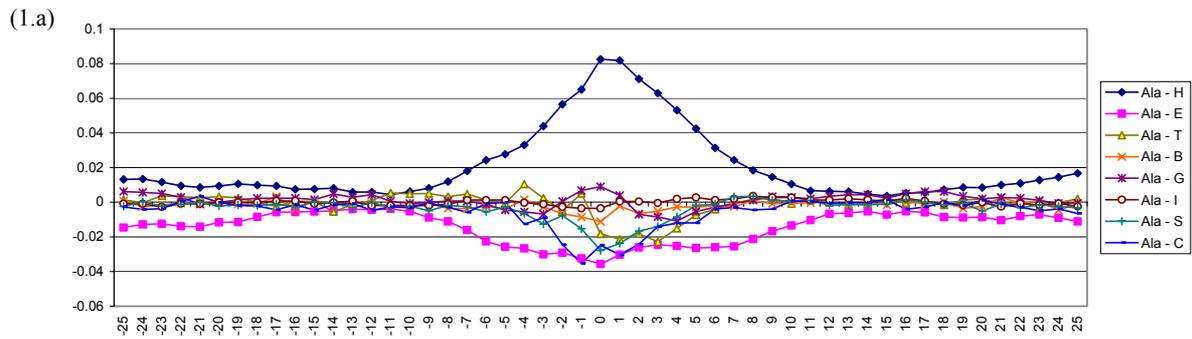

(1.b)
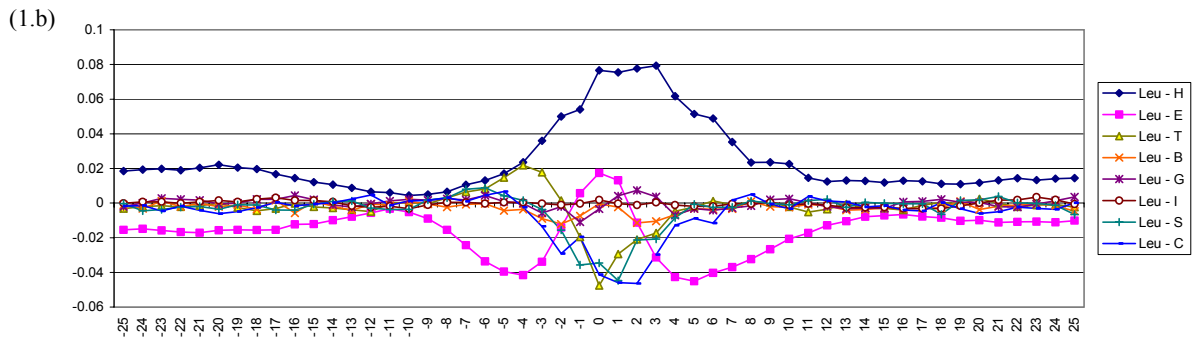

(1.c)
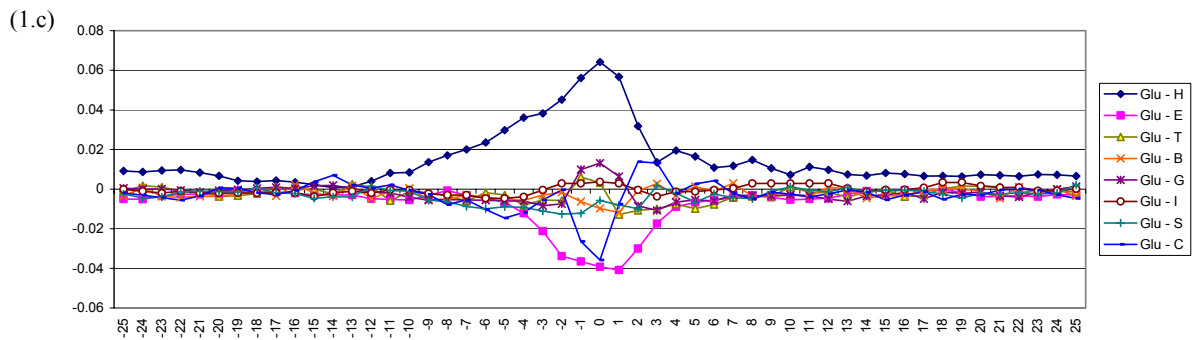

(1.d)
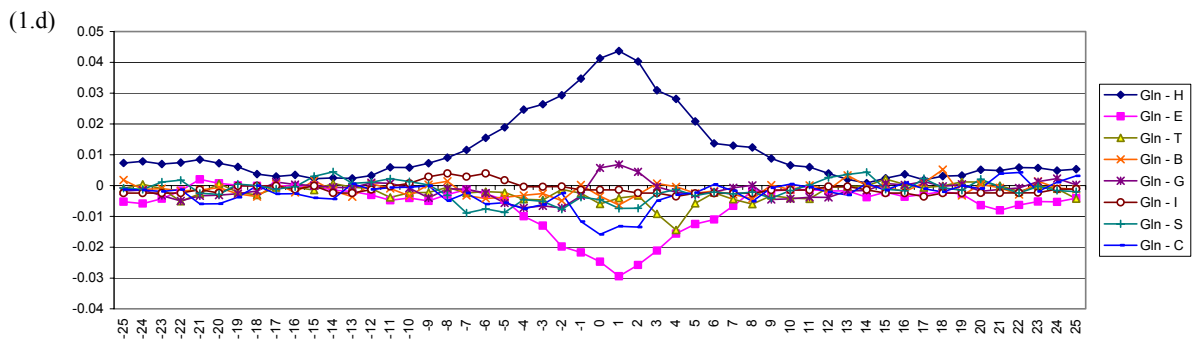

(1.e)
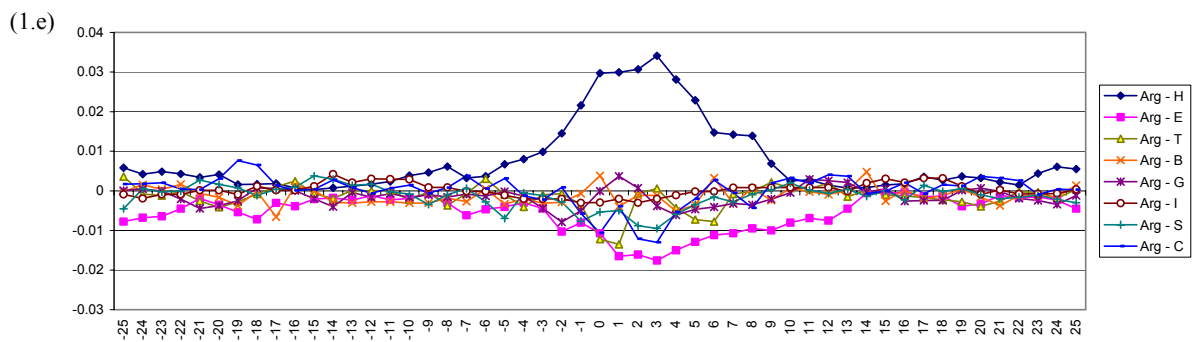

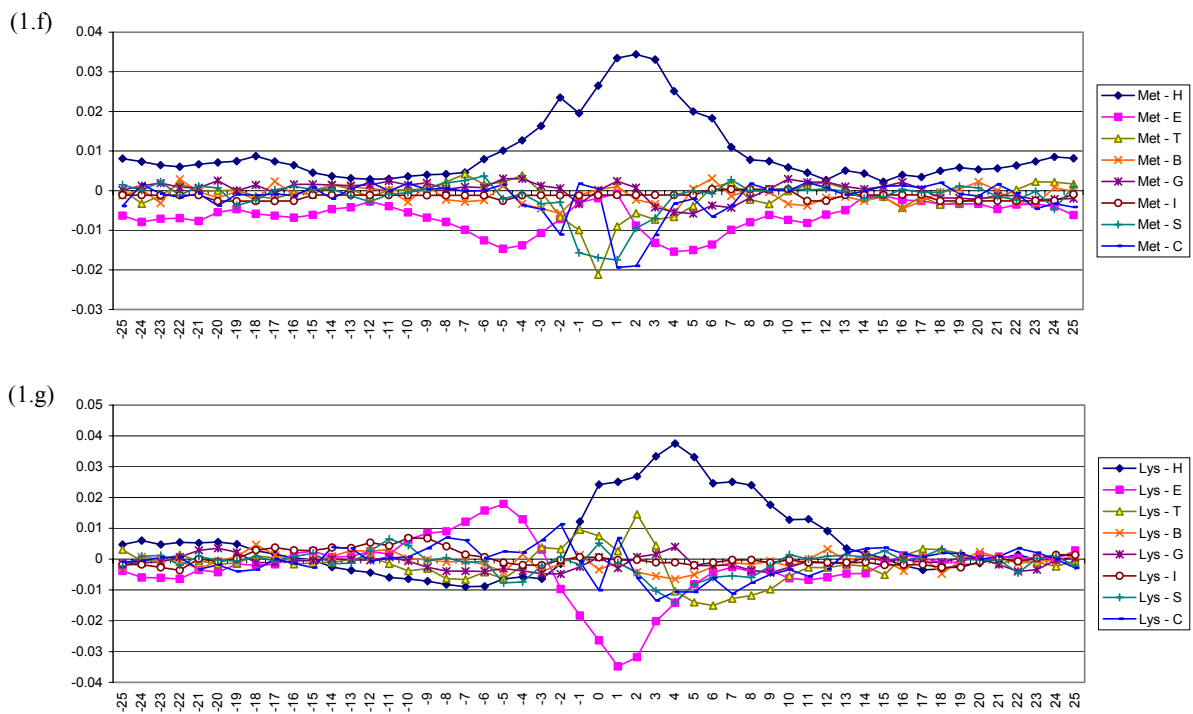

**Figure 1. Correlations of α-helix admirers with secondary structure types.** Each diagram contains eight graphs representing the correlation of an amino acid and the eight secondary structure types. The x-axis is labeled by $\tau$, the offset of the amino acid relative to the position of the considered secondary structure type. The positive (negative) values of $\tau$ correspond to amino acid positions following (preceding) the secondary structure type position (Equation 1). The y-axis is labeled by the correlation value. The range of $\tau$ is same in all diagrams, but the range of correlations (y-axis) varies significantly. Correlations for each of α-helix admirers are presented in a separate diagram: (a) Alanine, (b) Leucine, (c) Glutamic acid, (d) Glutamine, (e) Arginine, (f) Methionine and (g) Lysine.

The amino acids from this group show similar behavior with respect to other secondary structures, but there are also differences in their behavior. Amino acids *Ala*, *Glu*, *Gln*, *Arg* and *Lys* dislike strands and coils. At the contrary, *Leu* is a unique amino acid in this group that tends to build strands. It prefers short strands but obstructs both longer strands and coils. *Met* is relatively neutral to appearance in strands, 3-helices and coils. *Ala*, *Leu* and *Met* obstruct the formation of turns and bends. *Glu* supports the formation of short 3-helices.

**Strand admirers.** The amino acids from the second group (*Val*, *Ile*, *Tyr*, *Phe*, *Thr* and *Trp*) prefer strands. *Thr* is unique among strand admirers and among all amino acids because it has almost the same correlation value with strands and with coils. Based only on the correlation value for coils, it could be also classified as turn and bend admirer. Still, because of large negative correlation value for turns, we put it among strand admirers. However, it differs from other members of the group, what is obvious from Figure 2.

Results of Chou and Fasman (17) that *Val* and *Ile* prefer strands, and results of Gibrat *et al.* (50) that *Val*, *Ile*, *Tyr* and *Trp* prefer strands are in agreement with our results. Interestingly, in these previous results *Met* (17) and *Cis* (17, 50) were among the strongest β-sheet formers, while our data show that *Met* has slightly negative and *Cis* only a small positive correlation.

Though helix admirer, *Leu* is also positively correlated with strands, as already mentioned.

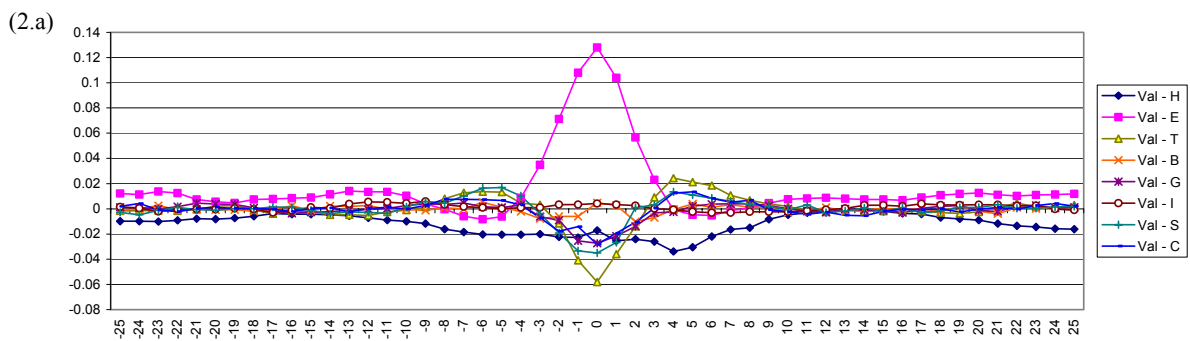

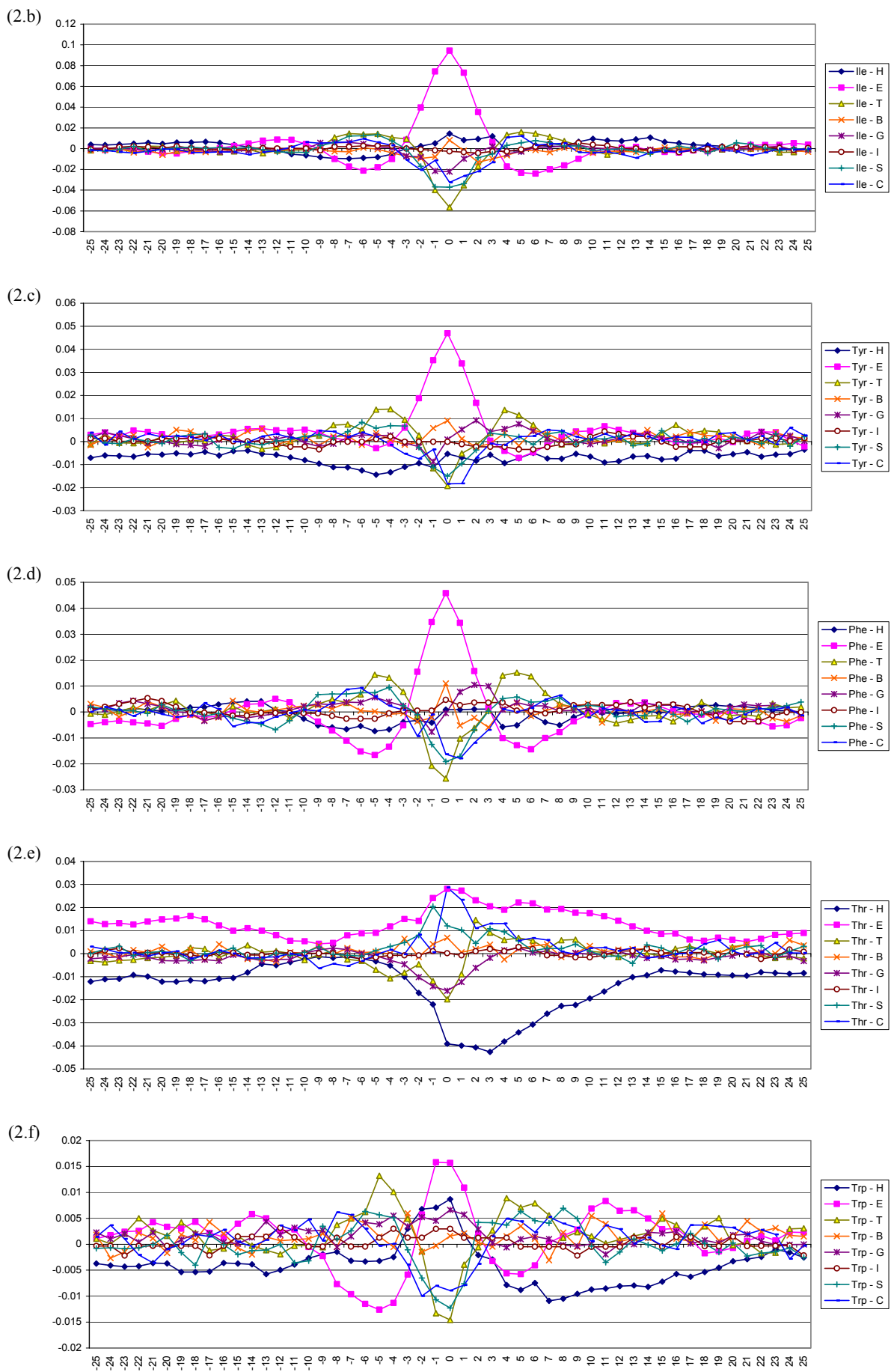

**Figure 2. Correlations of strand admirers with secondary structure types.** See Figure 1. for detailed description. Correlations for each of strand admirers are presented in a separate diagram: (a) Valine, (b) Isoleucine, (c) Tyrosine, (d) Phenylalanine, (e) Threonine and (f) Tryptophan.

Most of the strand admirers are weakly correlated with α-helices. *Val* obstructs α-helices while *Thr* strongly obstructs α-helices. All strand admirers obstruct the formation of turns and bends, except *Thr*, that supports the formation of bends. As it was mentioned, *Thr* is also unique among strand admirers because it supports the formation of coils. *Thr* and *Tyr* are positively correlated with β–bridges.

*Val* and *Ile* obstruct while *Trp* supports the formation of 3-helices.

**Turn and bend admirers.** The amino acids from the third group (*Gly*, *Asn*, *Pro*, *Asp* and *Ser*) show preference to build bends or turns or coils. This is good agreement with results of Chou and Fasman who found that *Pro*, *Gly*, *Asn* and *Ser* are the most frequent coil residues (17) and with Gibrat *et al*. (50), who added *Asp* to the set.

Turn and bend admirers, *Gly*, *Asn*, *Pro*, *Asp* and *Ser*, have quite large positive correlation values for bends, turns, 3-helices and coils. Only *Gly* has negative value for 3-helices, *Asn* very small value for 3-helices, and *Ser* for turns. All amino acids in this group occur rarely in α-helices and strands. *Gly* has very high tendency to build turns and appears very often at their end. *Pro* tends to initiate turns. *Pro* tends to appear in terminating parts of bends and coils. *Pro, Asp* and *Ser* support the formation of 3-helices.

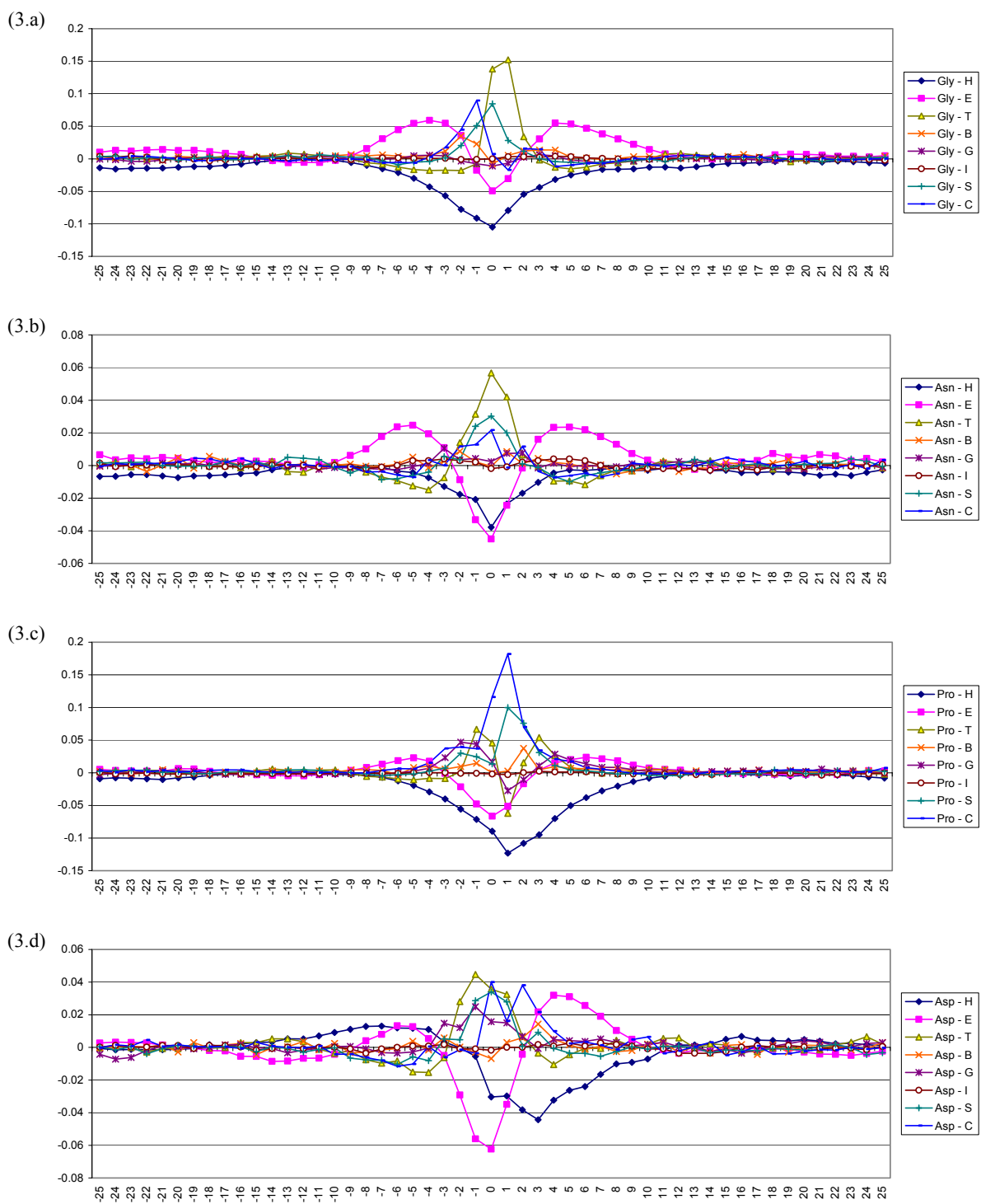

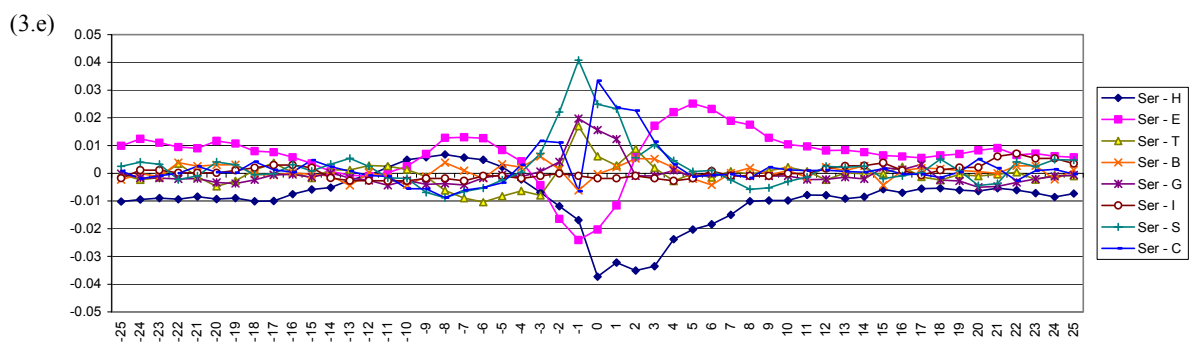

**Figure 3. Correlations of turn and bend admirers with secondary structure types.** See Figure 1. for detailed description. Correlations for each of turn and bend admirers are presented in a separate diagram: (a) Glycine, (b) Asparagine, (c) Proline, (d) Aspartic acid and (e) Serine.

*Cys* and *His*. The remaining two amino acids *Cys* and *His* are relatively weakly correlated with all secondary structure types. The correlation values suggest statistical significance, but compared to other amino acids these values are substantially lower. They do not show clear preference to build any secondary structure and do not show large negative correlation values for any of secondary structure types. With small correlation coeficients *Cys* tends to build strands, while *His* has negative correlation with α-helices. It is interesting that results of Chou and Fasman show large preference of *His* towards α-helices, and preference of *Cis* towards β-sheet structures (17).

*Influence Of Amino Acids Properties*

**Polarity and size of amino acids.** Amino acids tendencies to take part in certain secondary structure can be connected with physicochemical properties of amino acids. It is known that in strands there are predominantly hydrophobic amino acids (17).

It was recently shown that propensities of amino acids for certain position in helix depend on physicochemical properties (27). Polar and nonpolar amino acids show different phase distribution – they usually appear at different positions in helices. Long polar (*Glu*, *Gln*, *Arg*, *Lys*) and short polar (*Asn*, *Asp*, *Ser*) amino acids have the same phase distribution. Hydrophobic aromatic (*Phe*, *Tyr*, *Trp*) and hydrophobic aliphatic (*Leu*, *Met*, *Val*, *Ile*) have the same phase distribution that is opposite to phase distribution of polar amino acids (27). Only five amino acids fail to follow this pattern, *Gly*, *Ala*, *Thr*, *Pro*, and *His*. *Gly* and *Ala* have very small side chains; *Thr* has also small side chain and has intermediate polarity. *Pro* occurs very rarely in α-helices. *His* do not cluster either with polar or nonpolar because of its pKa that is near neutral pH (27).

Classification of amino acids as long polar (*Glu*, *Gln*, *Arg*, *Lys*), short polar (*Asn*, *Asp*, *Ser*), hydrophobic aromatic (*Phe*, *Tyr*, *Trp*), and hydrophobic aliphatic (*Leu*, *Met*, *Val*, *Ile*) could be connected with our results. All long polar amino acids are α-helix admirers, all aromatic are strand admirers, while all short polar are turn and bends admirers. However, aliphatic amino acids do not belong to one group of admirers (Figures 1-4, Table I), some of them are among α-helix admirers, and some of them are strand admirers. Looking at the structures of aliphatic amino acids it can be noticed that amino acids with branch at Cβ atom are among strand admirers (*Val* and *Ile*), while amino acids without branching on Cβ atom are α-helix admirers (*Ala*, *Leu*, *Met*). Some more general rules connecting structural properties of amino acids with our classification can be noticed, and we will discuss them latter.

As in the case of position in α-helices (27) *Thr*, *Gly* and *Pro* are exceptions. However, *Ala* is not an exception, because it is α-helix admirer, like *Leu* and *Met*.

Hence, there is connection of polarity and size of amino acids with our groups (Table I).

All strand admirers are hydrophobic, with exception of *Thr* that is slightly polar. The tendency of nonpolar amino acids for strand structures is known for long time (17, 50). It is also supported by the fact that all polar amino acids, that belong to α-helix admirers and turn and bend admirers, have negative correlation with strand structures, while hydrophobic *Leu* has positive correlation. This is in agreement with the finding that proteins with increased hidrophobicity are less stable against misfolding (51).

Among α-helix admirers there are hydrophobic and long polar amino acids. The tendency of α-helix for hydrophobic amino acid is also supported by the fact that hydrophobic aromatic amino acids have positive (*Phe*, *Trp*), or slightly negative (*Tyr*) correlation with α-helices, while all short polar amino acids show negative correlation.

All turn and bend admirers are small polar amino acids as well as *Gly* and *Pro*. The tendency of small polar amino acids to build turns, bends, and coils is also in agreement with positive correlation of *Thr* (that is classified as strand admirer) with bends and coils (Table I). *Gly* and *Pro* are only exceptions in this group, since they are not polar.

**Structural properties of amino acids.** Groups of amino acids given in Figures 1-4 and in Table I are formed based on our results about preference of amino acid to be part of certain secondary structure. However, we realized that amino acids that are in the same group have the same structural properties. This means that certain amino acids structural properties are attuned to certain secondary structure types.

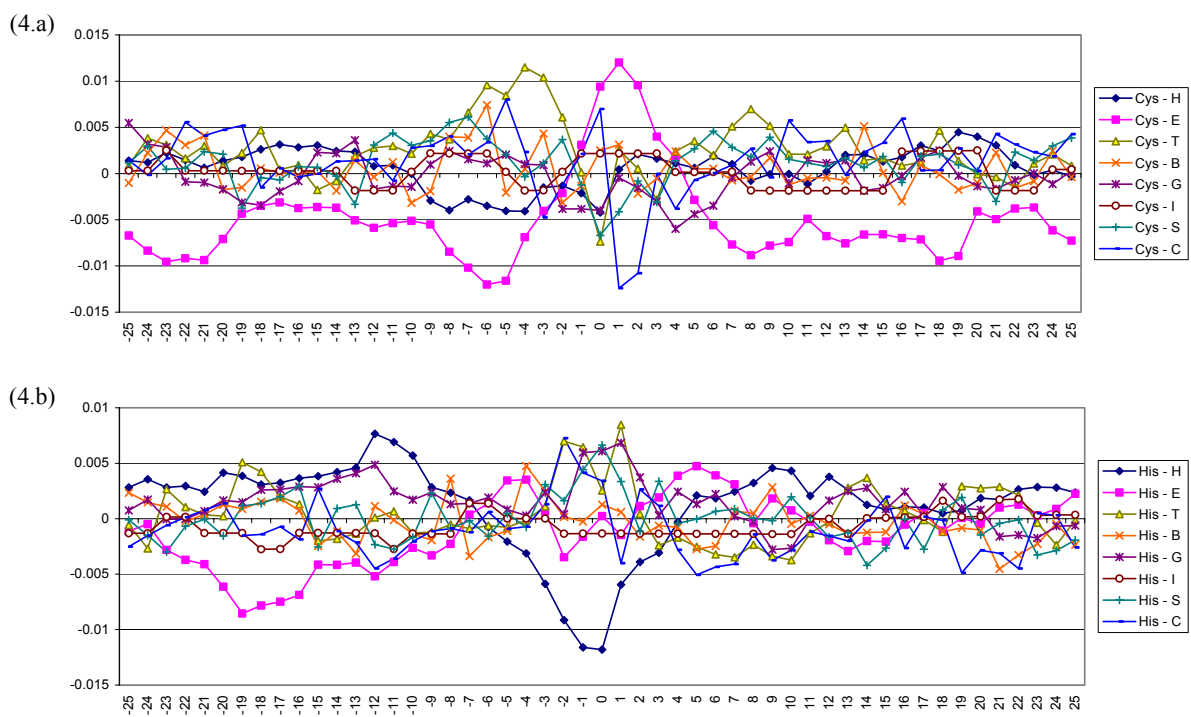

**Figure 4. Correlations of other amino acids with secondary structure types.** See Figure 1. for detailed description. Correlations for Cysteine (a) and Histidine (b) are presented in separate diagrams.

All α-helix admirers have common structural properties, and all amino acids with these structural properties are α-helix admirers. In α-helix admirers there are two hydrogens and one carbon (in *Ala* there are three hydrogens) and there is no branching on Cβ atom; Cγ atom is aliphatic (sp$^3$ hybridization) and there are no heteroatoms – only hydrogen and carbon atoms are on Cγ. The only exception is *Met* with sulfur on Cγ atom. *Met* is probably among α-helix admirers because sulfur in this case does not bring polarity. Hence, we can say that there are nonpolar heteroatoms on Cγ atom, and classify *Met* as α-helix admirer. There could be polar groups, or polar heteroatoms in structures of α-helix admirers, but further than Cγ atom.

All aromatic and all amino acids with branching on Cβ atom are strand admirers. *Thr* has branching on Cβ atom and it is among strand admirers, although it is polar amino acid and the rest are nonpolar.

All turn and bend admirers have polar heteroatoms on Cβ or Cγ atoms. *Pro* is also among turn and bend admirers with its unusual structure. However, it also has polar heteroatom, N, on Cβ.

The example of *Thr* is extremely interesting. It has branching on Cβ, and also polar heteroatom on Cβ. Hence it has structural properties of both strand admirers and turn and bend admirers. At the same time *Thr* indeed has almost the same correlation coefficient for strands and coils. This shows clearly that structural properties are connected very closely with the type of secondary structure forming by certain amino acid.

*His* and *Cis,* showing no preference for any of secondary structures, also do have structures quite different than all other amino acids. *His* has polar heteroatom on Cβ and at the same time that polar atom is part of the aromatic ring. *Cis* has not very polar SH group on Cβ atom, hence it differs from turn and bend admirers. Another difference is the fact that SH groups can make disulfide bridges, making this amino acid quite different from the others, and explaining why it do not belong to any of previous groups.

Based on these observations, it seem that tendency of amino acid to take part in certain secondary structure type is not defined by polarity or hidrophobicity of amino acid (although there is some connection), but that the crucial property of amino acid is the structure on Cβ or Cγ atoms – branching and polar heteroatoms. In other words it seems that the part of the amino acid that is close to the backbone determines the type of secondary structure.

**Amino Acids And Secondary Structure Types At Distant Positions**

Diagrams in Figures 1-4 show correlation of amino acid and the secondary structure at the offset $\tau$ varying from –25 to 25. It is seen that the range of the offset $\tau$ values, for which the correlation coefficients are significant (Equation 4), varies for different amino acids, and different secondary structures. If we take 0.004 (see Method/Data Sets) as the threshold correlation value, then the average range with significant correlations is [-9, 10]. In some cases, significant correlation values are obtained for $\tau$ beyond that range. For example, the amino acids that are helix admirers often have significant correlations with secondary structure positioned separated by 10-12 amino acids. In papers (39, 47, 50) a smaller range [-8,8] is considered to be significant. We assume that larger data set and different computational model provide a wider range of significant values.

*Propensities Of Amino Acids*

**α-helix admirers.** These amino acids support the appearance of α-helices in the vicinity, but there are differences in their appearance in different parts of helices. That is in agreement with previous results showing that certain amino acids have preference to be in the beginning or end of α-helix (27, 17, 47).

While *Ala* induces α-helices in its vicinity almost equally in both directions, the other amino acids from this group have asymmetrical distribution of the α-helices support. The asymmetry of *Gln*'s correlation is not large. The correlation distribution for amino acids *Leu*, *Arg*, *Met* and *Lys* with α-helices is shifted towards positive values of $\tau$, with the largest shift for *Lys*. At the contrary, the correlation for *Glu* is very strongly shifted towards negative values of $\tau$. That leads to the conclusion that *Ala* and *Gln* are evenly distributed in, while *Leu*, *Arg*, *Met* and *Lys* tend to be closer to the end of helices. *Glu* tends to be very close to the beginning of α-helices, agreeing very well with (27).

Amino acids *Ala*, *Glu*, *Gln*, *Arg* dislike strands and coils. All α-helix admirers, except *Lys*, obstruct the formation of strands in their neighborhood. *Lys* obstructs the formation of strands in its and preceding positions, but strongly supports their formation in subsequent positions, which can be explained by its higher α-helix propensity at the helix end. As mentioned before, *Leu* is a unique amino acid in this group that tends to build strands, however, it is interesting that *Leu* has strong negative correlation with strands at distance 5-6 in both directions. *Met* is relatively neutral to appearance in strands, but it has negative correlation with strands in its neighborhood.

*Lys* supports turns formation in its immediate neighborhood. *Met* obstructs the formation of coils in the preceding positions. *Glu* strongly supports the formation of coils in the preceding positions.

**Strand admirers.** It is interesting that most of amino acids from the strand admirers group (*Val*, *Ile*, *Tyr*, *Phe*, *Thr* and *Trp*), do not have positive correlation with strands at the distance. Amino acids *Ile*, *Phe* have quite negative correlation with strands at distance of 4 to 8, while *Val*, and *Trp* have less pronounced but still negative correlation at the same distance. *Trp* has stronger negative correlation for negative values of $\tau$. It is interesting that *Val* again has positive correlation with strands at distances larger than 8 positions. *Tyr* is relatively neutral to strands in its vicinity. *Thr* differs from other members of the group, having positive, but small correlation with strands for all $\tau$ values.

Many of strand admirers have negative correlation with distant α-helix structures. *Thr* and *Val* have negative correlation with α-helices for positive values of $\tau$. *Ile* shows small positive correlation for almost all positive values of $\tau$. There is also uniform small negative correlation between *Tyr* and α-helices for both positive and negative values of $\tau$.

Most of strand admirers have positive correlation with distant turns. *Thr* is an exception: it supports the formation of turns in the preceding, but has negative correlation with them in the subsequent positions.

All strand admirers, except *Thr*, obstruct the formation of bends. *Thr* weakly supports the formation of bends, mostly in positions between –2 and 5.

*Thr*, *Tyr*, *Phe* and *Ile* are positively correlated with β–bridges. *Val*, *Ile* and *Thr* obstruct the formation of 3-helices, while *Phe* and *Tyr* support them at both its and immediately preceding positions.

*Thr* supports the formation of coils at its and precedent positions. *Ile*, *Val*, *Tyr* and *Phe* obstruct coils. *Val* tends to support coils in its vicinity.

**Turn and bend admirers.** The amino acids from the third group (*Gly*, *Asn*, *Pro*, *Asp* and *Ser*), turn and bend admirers, are frequent in the vicinity of strands, but have negative correlation with α-helices in their vicinity, except *Asp* for $\tau < 0$. It supports α-helices in the subsequent positions. Most of them have positive correlation with bends, turns and coils in their close vicinity, but often asymmetrically.

*Asp* supports coils at it's and the preceding positions ($\tau \geq 0$). *Gly* tends to build turns and appears very often at their end. It supports the formation of coils at the subsequent position ($\tau = -1$). *Pro* tends to initiate turns and to support turns in its vicinity, but it strongly obstructs the formation of turns at $\tau = +1$. *Pro* tends to appear in terminating parts of bends and coils. *Asp* and *Ser* support the formation of 3-helices.

***Cys* and *His*.** As it was mentioned the remaining two amino acids *Cys* and *His* are relatively weakly correlated with secondary structures, and it is also the case for their vicinity. With low correlation value *Cys* has negative correlation with strands for $\tau$ values below –4 and above +5, while *His* has small negative correlation with α-helices for $\tau$ between -3 and 1.

As discussed above, the amino acids belonging to the same group often tend to express similar behavior in respect to certain secondary structure type, not only to the one they admire, but also to the others. It is interesting to compare the diagrams to those of Robson and Suzuki (47). There are substantial similarities for most of amino acids (*Ala*, *Leu*, *Glu*, *Lys*, *Val*, *Ile*, *Gly*, *Asn*, *Pro*, *Asp* and *Ser*), but there are substantial differences also (*Gln*, *Arg*, *Phe*, *Trp* and *His*).

*Influence Of Amino Acids Properties*

It is interesting to connect shape of the diagrams with structural properties of amino acids. The large asymmetry for α-helices is observed for *Glu* and *Asp* that have carboxyl groups. *Arg* and *Lys*, amino acids with amino groups, also have asymmetric diagrams for α-helices and strands. Hence, polarity and charge of amino acids, as well as possibility to be hydrogen bond acceptor or donor, is connected with the similar influence of amino acids on the distant secondary structure. Diagrams for *Ser* and *Thr* show asymmetry and similar behavior of these two amino acids in respect to α-helix and strand structures. It is again connected with their similar structure, OH group on Cβ atom.

The asymmetry is the most pronounced for α-helix structures. Obviously it is connected with hydrogen bonds that exist in α-helices.

From the data presented in Figures 1-4 and in Table I it seems that polarity, charge and capability for hydrogen bonding have more influence on the distant secondary structure, while structural properties of Cβ or Cγ atoms have influence on the secondary structure at the position of amino acid (τ near zero). Hence, atoms close to the backbone define secondary structure at the position of amino acid, while hydrogen bonding and other noncovalent interactions between amino acids have influence on distance.

## Conclusions

The calculated correlations of amino acids with secondary structure types enable to determine amino acid tendency to participate in certain secondary structure type. Results confirm that there is a significant dependence of secondary structure type on amino acid, not only on the corresponding position, but also on amino acids in nearby positions. It confirms that the secondary structure prediction should equally consider preceding and subsequent amino acids. In most cases, the type of the secondary structure depends on 9 preceding and 10 subsequent amino acids.

Results show that most of amino acids have clear preference to participate in certain secondary structure type. Based on it amino acids are classified in four groups: α-helix admirers, strand admirers, turn and bend admirers and the others that do not show preference for any secondary structure. The amino acids from the same group often show similar behavior to certain secondary structure type, not only to the one they admire, but also to the others.

Analyzing physicochemical and structural properties of amino acids show that amino acids from the same group have similar physicochemical and the same structural characteristics. Nonpolar aliphatic without branching on Cβ atom and long polar amino acids are α-helix admirers. Strand admirers are aromatic and aliphatic amino acids with branching on Cβ atom, while turn and bend admirers are small polar amino acids. The common structural properties of α-helix admirers are: no polar atoms on Cβ and Cγ atoms, no branching on Cβ, and aliphatic (sp$^3$) Cγ atom. All amino acids that have aromatic groups or branching on Cβ atom are strand admirers. All turn and bend admirers have polar heteroatom on Cβ or Cγ atoms, or do not have Cβ atom. Hence, based only on structure amino acid can be classified to certain group.

Diagrams that show correlation of amino acid with secondary structure at neighboring positions, for offset τ values from −25 to 25, reveal that the distance range of significant correlation is usually between -9 and 10. Shape of the diagrams is connected to structural properties of amino acids. The largest asymmetry is observed for polar amino acids, and amino acids that can make hydrogen bonds. These results indicate that polarity and capability for hydrogen bonding have influence on the secondary structure at some distance. However, polarity and hydrogen bonding do not have crucial influence on preference for certain secondary structure type. Our results suggest that amino acid preference for secondary structure is caused by structural properties of Cβ or Cγ atoms.

## References

[1] Rost B. Protein structure prediction in 1D, 2D, and 3D. In: von Rague-Schleyer P, Allinger NL, CClark T, Gasteiger J, Kollman PA, Schaefer HF. Encyclopedia of Computational Chemistry. Sussex: John Wiley; 1998. p 2242-2255.

[2] Bowie JU, Luthy R, Eisenberg D. A method to identify protein sequences that fold into a known three-dimensional structure. Science 1991; 253:164-170.

[3] Fischer D, Eisenberg D. Protein fold recognition using sequence-derived predictions. Protein Science 1996; 5:947-955.

[4] Maiorov VN, Crippen GM. Contact potential that recognizes the correct folding of globular proteins. J Mol Biol 1992; 227:876-888.

[5] Koretke KK, Luthey-Schulten L, Wolynes PG. Self-consistently optimized energy functions for protein structure prediction by molecular dynamics. Proc Natl Acad Sci USA 1998; 95:2932-2937.

[6] Bastolla U, Vendruscolo M, Knapp EW. A statistical mechanical method to optimize energy functions for protein folding. Proc Natl Acad Sci USA. 2000; 97:3977-3981.

[7] Solis AD, Rackovsky S. On the use of secondary structure in protein structure prediction: a bioinformatics analysis. Polymer 2004; 45:525-546.

[8] Ortiz AR, Kolinski A. Rotkiewicz P, Ilkowsky B, Skolnick J. Ab initio folding of proteins using restraints derived from evolutionary information. Proteins 1999; Suppl 3:177-185.

[9] Eyrich VA, Standley DM, Felts AK, Friesner RA. Protein tertiary structure prediction using a branch and bound algorithm. Proteins 1999; 35:41-57.

[10] Eyrich VA, Standley DM, Friesner RA. Prediction of protein tertiary structure to low resolution: performance for a large and structurally diverse test set. J Mol Biol 1999; 288:725-742.

[11] Lomize AL, Pogozheva ID, Mosberg HI. Prediction of protein structure: the problem of fold multiplicity. Proteins 1999; Suppl 3:199-203.

[12] Chen CC, Singh JP, Altman RB. Using imperfect secondary structure predictions to improve molecular structure computations. Bioinformatics 1999; 15:53-65.

[13] Levitt M, Warshel A. Computer simulation of protein folding. Nature 1975; 253:694-698.

[14] Samudrala R, Xia Y, Huang E, Levitt M. Ab initio protein structure prediction using a combined hierarchical approach. Proteins 1999; Suppl 3:194-198.

[15] Samudrala R, Huang E, Koehl P, Levitt M. Constructing side chains on near-native main chains for ab initio protein structure prediction. Protein Eng 2000; 13:453-457.

[16] Kelley LA, MacCallum RM, Sternberg MJE. Enhanced Genome Annotation Using Structural Profiles in the Program 3D-PSSM. J Mol Biol 2000; 299:499-520.

[17] Chou PY, Fasman GD. Conformational parameters for amino acids in helical, beta-sheet and random coil regions calculated from proteins. Biochem 1974; 13(2):211-222.

[18] Chou PY, Fasman GD. Prediction of the secondary structure of proteins from their amino acid sequence. Advan Enzymol Relat Areas Mol Biol 1978; 47:45–148.

[19] O'Neil KT, DeGrado WF. A thermodynamic scale for the helix-forming tendencies of the commonly occurring amino acids. Science 1990; 250:646–651.

[20] Padmanabhan S, Marqusee S, Ridgeway T, Laue TM, Baldwin RL. Relative helix-forming tendencies of nonpolar amino acids. Nature 1990; 344:268–270.

[21] Kim CA, Berg JM. Thermodynamic β-sheet propensities measured using a zinc-finger host peptide. Nature 1993; 362:267–270.

[22] Minor DL, Kim PS. Measurement of the β-sheet-forming propensities of amino acids. Nature 1994; 367:660–663.

[23] Smith CK, Withka JM, Regan L. A thermodynamic scale for the β-sheet forming tendencies of the amino acids. Biochemistry 1994; 33:5510–5517.

[24] Rost B. Review: Protein Secondary Structure Prediction Continues to Rise. J Struc Biol 2001; 134:204-218.

[25] Penel S, Hughes E, Doig AJ. Side-chain Structures in the First Turn of the α-Helix. J Mol Biol 1999; 287:127-143.

[26] Petukhov M, Muñoz V, Yumoto N, Yoshikawa S, Serrano L. Position Dependence of Non-polar Amino Acid Intrinsic Helical Propensities. J Mol Biol 1998; 278:279-289.

[27] Engel DE, DeGrado WF. Amino Acid Propensities are Position-dependent Throughout the Length of Helices. J Mol Biol 2004; 337(5):1195-1205.

[28] Mandel-Gutfreund Y, Gregoret LM. On the significance of alternating patterns of polar and non-polar residues in β-strands, J Mol Biol 2002; 323(3):453-61.

[29] Minor DL, Kim PS. Context is a major determinant of β-sheet propensity. Nature 1994; 371:264–267.

[30] Dalal S, Balasubramanian S, Regan L. Protein alchemy: changing β-sheet into alpha-helix. Nature Struct Biol 1997; 4:548–552.

[31] Xiong H, Buckwalter BL, Shieh HM, Hecht MH. Periodicity of polar and nonpolar amino acids is the major determinant of secondary structure in self-assembling oligomeric peptides. Proc Natl Acad Sci USA 1995; 92:6349–6353.

[32] Baldwin RL, Rose GD. Is protein folding hierarchic? I. Local structure and peptide folding. Trends Biochem Sci 1999; 24:26–33.

[33] Baldwin RL, Rose GD. Is protein folding hierarchic? II. Folding intermediates and transition states. Trends Biochem Sci 1999; 24:77–83.

[34] Rost B. Prediction in 1D: Secondary structure, membrane helices and accessibility. In: Bourne PE, Weissig H, editors. Structural Bioinformatics. Hoboken: Wiley-Liss; 2003. p 559-587.

[35] Frishman D, Argos P. Incorporation of non-local interactions in protein secondary structure prediction from the amino acid sequence. Protein Eng 1996; 9(2):133–142.

[36] Frishman D, Argos P. Seventy-five percent accuracy in protein secondary structure prediction. Proteins 1997; 27:329–335.

[37] Salamov AA, Solovyev VV. Protein secondary structure prediction using local alignments. J Mol Biol 1997; 268:31–36.

[38] Gromiha MM, Selvaraj S. Inter-residue interactions in protein folding and stability. Progress in Biophysics & Molecular Biology 2004; 86(2):235-277

[39] Garnier J, Osguthorpe DJ, Robson B. Analysis and implications of simple methods for predicting the secondary structure of globular proteins. J Mol Biol 1978; 120:97-120.

[40] Chou PY, Fasman GD. Prediction of protein conformation. Biochem 1974; 13:222-245.

[41] Robson B. Analysis of code relating sequences to conformation in globular proteins. Theory and application of expected information. Biochem J 1974; 141(3):853-867.

[42] Berman HM, Westbrook J, Feng Z, Gilliland G, Bhat TN, Weissig H, Shindyalov IN, Bourne PE. The Protein Data Bank. Nucl Acids Res 2000; 28(1):235-242

[43] Zaric SD, Popovic DM, Knapp EW. Metal ligand aromatic cation-pi interactions in metalloproteins: ligands coordinated to metal interact with aromatic residues. Chemistry 2000; 6(21):3935-42.

[44] Zaric SD, Popovic DM, Knapp EW. Factors determining the orientation of axially coordinated imidazoles in heme proteins. Biochem 2001; 40(26):7914-28.

[45] Kabsch W, Sander C. Dictionary of protein secondary structure: pattern recognition of hydrogen-bonded and geometrical features. Biopolymers 1983; 22(12):2577-2637.

[46] Kloczkowski A, Ting KL, Jernigan RL, Garnier J. Combining the GOR V Algorithm With Evolutionary Information for Protein Secondary Structure Prediction From Amino Acid Sequence. Proteins 2002; 49:154-166.

[47] Robson B, Suzuki E. Conformational Properties of Amino Acid Residues in Globular Proteins. J Mol Biol 1976; 107:327-356.

[48] Samuels ML, Witmer JA. Statistics for the Life Sciences, 3rd ed. New Jersey, USA: Pearson Education; 2003.

[49] Hobohm U, Sander C. Enlarged representative set of protein structures. Protein Sci 1994; 3:522-524.

[50] Gibrat JF, Garnier J, Robson B. Further Developments of Protein Secondary Structure Prediction Using Information Theory: New Parameters and Consideration of Residue Pairs. J Mol Biol 1987; 198:425-443.

[51] Bastolla U, Moya A, Viguera E, van Ham RCHJ. Genomic Determinants of Protein Folding Thermodynamics in Prokaryotic Organisms. J Mol Biol 2004; 343:1451-1466.